\documentclass[twocolumn,showpacs,preprintnumbers,amsmath,amssymb]{revtex4}

\usepackage{graphicx}
\usepackage{dcolumn}
\usepackage{bm}
\newcommand{\bfr}{{\bf r}}


\begin{document}

\title{Vortex waves in trapped Bose-Einstein condensates}
\author{T. P. Simula, T. Mizushima, and K. Machida}
\affiliation{Department of Physics, Okayama University, Okayama 700-8530, Japan}

\begin{abstract}
We have theoretically studied vortex waves of Bose-Einstein condensates in elongated harmonic traps. Our focus is on the axisymmetric varicose waves and helical Kelvin waves of singly quantized vortex lines. Growth and decay dynamics of both types of vortex waves are discussed. We propose a method to experimentally create these vortex waves on demand. 
\end{abstract}

\maketitle
\section{Introduction}
Substance whirling around a line in space forms a vortex. Eddies in a flowing water and whirlwinds in the atmosphere are perhaps the most common examples of vortices encountered in nature, although such structures exist in nearly every imaginable medium including laser light and superconductors. Vortices are objects of fundamental importance in fluid dynamics and they are inherent to turbulent flows featuring normal mode excitations analogous to vibrating classical strings \cite{FetterWalecka}. Lord Kelvin (Sir W. Thomson) derived a dispersion relation for a class of normal modes of classical hydrodynamic vortices \cite{Thomson1880a}. A particular type of vortex excitations in which the vortex twists around its equilibrium position forming a helical structure have since become known as Kelvin waves \cite{Donnelly1991a}. Such chiral vortex vibrations are an elemental property of vortices ranging from tiny whirlpools in creeks to tornadoes of atmospheric scales. Another type of axially propagating excitation mode---coined a varicose wave---may exist in fluids confined to cylindrical geometry. These radially pulsating modes have the varicose shape, well known in the context of blood flow through veins in the cardiovascular systems \cite{Ellis1999a}. 

Superfluids are distinguished from classical fluids by their ability of frictionless flow \cite{LeggettBook}. Rotational motion of superfluids is supported by quantized vortices \cite{Donnelly1991a}. Such quantum vortices possess many properties akin to their classical counterparts---including Kelvin waves \cite{Hall1958,Andronikashvili1960a,Pitaevskii1961a,Ashton1979,Sonin1987a,Donnelly1991a,Epstein1992a,Bretin2003a,Fetter2004a,Fetter2008a}. Superfluid Kelvin waves were first experimentally introduced in the context of liquid helium \cite{Hall1958,Andronikashvili1960a,Ashton1979, Donnelly1991a}. Later the discovery of Bose-Einstein condensates of dilute atomic gases \cite{Anderson1995a,Davis1995a, Bradley1995a} facilitated the creation and direct imaging of quantized vortices in highly controllable superfluid systems \cite{Matthews1999b,Madison2000a}. In particular, Kelvin waves on an isolated quantized vortex line were explicitly excited and imaged for the first time by Bretin {\emph et al.} \cite{Bretin2003a}. Their excitation method was described in terms of Beliaev decay of the quadrupole mode into a pair of Kelvin waves \cite{Bretin2003a,Mizushima2003a}. Further insight for this process was recently obtained by direct dynamical simulations \cite{Simula2008b}. Hydrodynamically it may be understood in terms of a parametric resonance between the quadrupole mode and a pair of Kelvin modes of opposite axial momenta, which induces an elliptical instability to the initially straight vortex line \cite{Kerswell2002a,Simula2008b}. The dynamics of superfluid Kelvin waves is crucial to understanding the details of quantum turbulence \cite{Vinen2006a,Halperin2008a}. It is also conceivable that the quantum gases would support varicose waves in which an axisymmetric pulsating perturbation propagates along the vortex line causing periodic modulation of the vortex core diameter.  

In this paper, we mainly focus on two types of vortex waves in trapped Bose-Einstein condensed gases. In the Sec. II we describe our numerical methods and in the Sec. III we first extend our previous results \cite{Simula2008b} on Kelvin waves followed by an introduction to varicose wave excitations.  Finally, the Sec. IV is devoted to discussion.

\section{Methods}
We consider a Bose-Einstein condensate composed of $^{87}$Rb particles trapped in a harmonic potential whose transverse and axial frequencies are $\omega_x=\omega_y=\omega_\perp=2\pi\times 98.5$ Hz and $\omega_z=2\pi\times11.8$ Hz, respectively, as in the experiment of  Bretin \emph{et al.} \cite{Bretin2003a}. The number of particles in the condensate is $1.3\times10^5$ unless otherwise stated. The ground state and its dynamics are accurately described by the time-dependent Gross-Pitaevskii equation \cite{PethickSmith,StringariBook}
\begin{equation}
i\hbar \partial_t\psi(\bfr,t)=\left( H_0+ V_{\rm pert}(\bfr,t)   + g|\psi(\bfr,t)|^2 \right) \psi(\bfr,t),
\label{GPE}
\end{equation}
where the single-particle quantum harmonic oscillator operator is given by
\begin{equation}
H_0=-\frac{\hbar^2\nabla^2}{2m}  + \frac{1}{2}m(\omega_x^2x^2 + \omega_y^2y^2+ \omega_z^2z^2),
\label{SHO}
\end{equation}
and the last term in  Eq.(\ref{GPE}) accounts for the contact interaction of strength $g$ between particles. The time-dependent potential $V_{\rm pert}(\bfr,t)$ may be engineered to resonantly excite various normal modes by perturbing the ground state. Such collective excitation modes are also obtained within the linear response theory by solving the coupled Bogoliubov-de Gennes eigenvalue equations \cite{PethickSmith,StringariBook}
\begin{eqnarray}
\mathcal{L}u_q(\bfr) + g\psi(\bfr)^2v_q(\bfr) &=&E_qu_q(\bfr) \nonumber \\
\mathcal{L}v_q(\bfr) + g\psi(\bfr)^{*2}u_q(\bfr)&=&-E_qv_q(\bfr),
\label{BDG}
\end{eqnarray}
where $u_q$ and $v_q$ are the usual quasiparticle wavefunctions corresponding to the excitations with eigenenergies $E_q$, and the single-particle operator
\begin{equation}
\mathcal{L}=H_0 +2g|\psi(\bfr,t)|^2-\mu
\label{LOP}
\end{equation}
contains the condensate chemical potential $\mu$.

We solve the time-dependent Gross-Pitaevskii equation by utilizing a parallel MPI implementation of a finite-element discrete variable representation method in a fully three dimensional cartesian coordinate system \cite{Schneider2006a,Simula2008a}. For finding solutions to the Bogoliubov-de Gennes equations, we deploy the cylindrical symmetry of the axisymmetric single vortex stationary state. As a result we have a good azimuthal quantum number $\ell$ thus reducing the effective computational effort by one spatial dimension. To diagonalize the matrix form of Eq.(\ref{BDG}), we use a hybrid basis consisting of radial Laguerre polynomials and finite-difference discretation in the axial direction. This yields a matrix which has a computationally beneficial band shape.

\section{Results}
By solving the eigenvalue problem for the axisymmetric single vortex initial state we obtain full information of the small-amplitude normal modes of our superfluid vortex state. Using dynamical simulations we can study the excitation and subsequent decay dynamics of vortex waves beyond linear stability analysis. Combining these two resources we are able to analyze excitation mechanisms, dispersion relations, and decay processes of the collective vortex excitations. In the following we separately focus on helical $\ell=-1$ Kelvin waves and axisymmetric $\ell=0$ varicose waves.

\subsection{Helical $\ell=-1$ Kelvin waves}

\emph{Physical characterization.---}Kelvin waves or their quanta---kelvons---are helical vortex waves propagating along the vortex line with an axial wave vector ${\bf k}_z$. The path ${\bf s_0}(\zeta,t)$ traced by the core of a vortex executing an idealized classical Kelvin wave motion of an amplitude $R$ may be curve parametrized as 
\begin{equation}
{\bf s_0}(\zeta,t)= R\cos(k_z \zeta+\omega t) {\bf\hat{e}_x} - R\sin(k_z \zeta+\omega t) {\bf\hat{e}_y}+\zeta {\bf\hat{e}_z},
\label{Kelvincurve}
\end{equation} 
where $t$ denotes time and $\omega$ is the rotation frequency of the line element about the $z$-axis. A particle following such trajectory has a negative helicity. One may use the Biot-Savart law to obtain the velocity field \cite{Donnelly1991a}
\begin{equation}
{\bf v}({\bf r})=\frac{\hbar}{ 2m}\int \frac{({\bf s}_0-{\bf r})\times d{\bf s}_0}{|{\bf s}_0-{\bf r}|^3}.
\label{BS}
\end{equation}
The divergence in the integral may be removed by introducing a cutoff $a_0=|{\bf r}-{\bf s}_0|$, which causes the internal structure of the vortex core to be neglected. Within the local induction approximation the self-induced fluid velocity in the vicinity of the vortex core is given by \cite{ArmsHama,Donnelly1991a}, 
\begin{equation}
{\bf v}_{\rm loc}({\bf r})=C {\bf s'}\times {\bf s''},
\label{BS}
\end{equation}
where $C$ is a system dependent constant and a prime denotes a derivative with respect to the curve parameter so that ${\bf s''}$ is the local curvature vector of the vortex and ${\bf s'}$ is tangent to the vortex line, pointing to the direction of the local vorticity vector. The Kelvin-Helmholz theorem guarantees the conservation of circulation which implies that a vortex tends to move with the fluid it is immersed in. Hence according to Eq.(\ref{BS}) the local motion of the vortex line element is in the direction determined by the binormal of the vorticity and the curvature vectors. In a uniform system this then implies that the Kelvin wave vortex helix must rotate in the direction opposite to the fluid flow around the unperturbed vortex.  

However, in the presence of external fields such as a harmonic trapping field, the local curvature ${\bf s''}$ may become significantly modified. A particularly elusive example is a singly quantized vortex precessing about the trap axis in a  harmonically trapped Bose-Einstein condensate \cite{Anderson2000a}. In such system the vortex is bent even in the pancake limit and the effective curvature vector points away from the trap centre and therefore the local self-induced velocity, which determines the direction of vortex precession, is in the same direction as the unperturbed condensate flow. The Kelvin modes correspond to a small amplitude motion of the vortex core and the corresponding mode densities are highly localized within the core volume of the vortex. 

\begin{figure}
\includegraphics[width=0.9\linewidth]{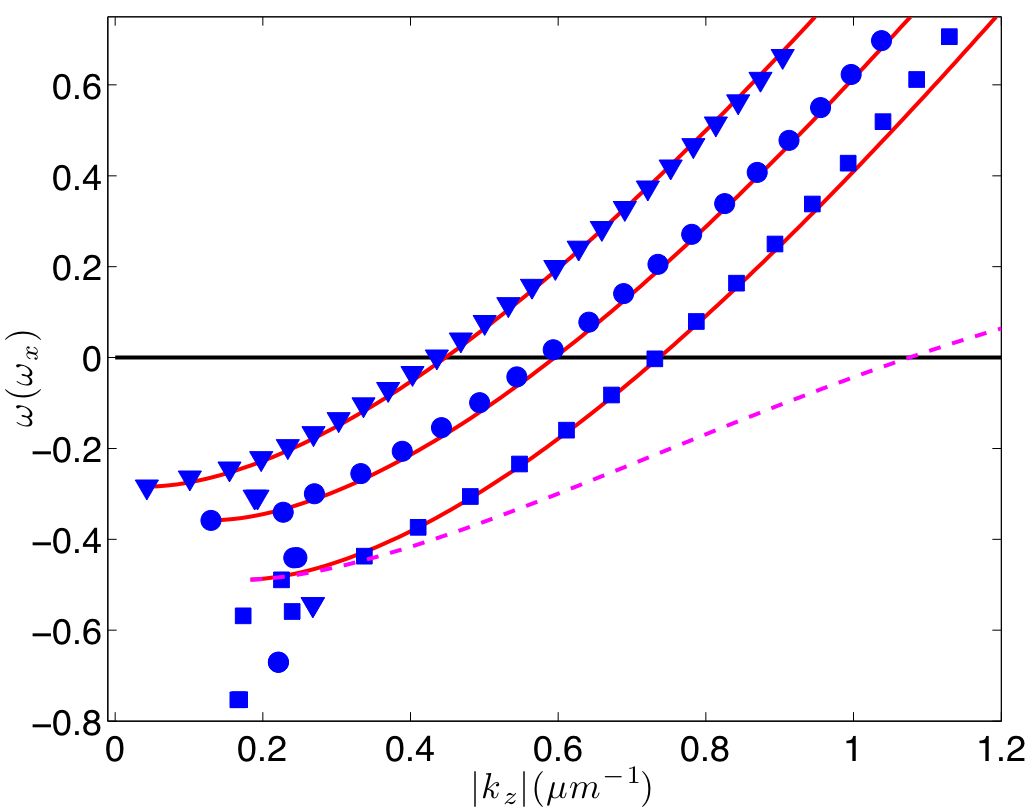}
\caption{Kelvon dispersion relation for different system sizes, $N=1\times10^4$ (squares), $5\times10^4$ (bullets), and $5\times10^5$ (triangles). The solid curves are plotted using the dispersion relation, Eq. (\ref{KDR}), with core parameter $r_c=0.13$ $\mu$m for all cases. The dashed line illustrates the sensitivity of the dispersion relation to the change of the core parameter as specified in the text.}
\label{Fkelvons}
\end{figure}

\emph{Dispersion relation.---}
The driving force of Kelvin wave motion involves velocity instead of acceleration and the resulting semiclassical long-wavelength, $|a_0k_z|\ll 1$, dispersion relation $\omega(k_z) = \hbar k_z^2  \ln \left( 1/|a_0 k_z| \right)/2m $ depends logarithmically on the wave vector $k_z$. In the case of an axisymmetric vortex line in a harmonically trapped Bose-Einstein condensate the kelvon dispersion relation is better described by \cite{Simula2008b}
\begin{equation}
\omega(k_0+k_z ) = \omega_0 +\frac{\hbar k^2_z}{2m}  \ln \left( \frac{1}{|r_c k_z|} \right) , \hspace*{2mm} |r_ck_z|\ll 1,
\label{KDR}
\end{equation}
where $r_c$ is a core parameter which is of the order of the vortex core size. The ``continuum" branch described by the above formula is cut off at low frequency $\omega_0<0$ and long-wavelength $k_0\sim2\pi/D_z$, where  $D_z$ is the Thomas-Fermi length of the condensate. In addition, the spectrum contains low-energy bending modes which lie below the continuum branch. From a qualitative point of view modes with $\omega<0$ rotate in the \emph{same direction} as the unperturbed condensate flow, whereas modes with $\omega>0$ rotate in the \emph{opposite direction}. Modes with zero frequency remain stationary in the laboratory frame. Figure \ref{Fkelvons} shows the computed kelvon dispersion relations for three different system sizes $N=1\times 10^4$ (squares), $N=5\times 10^4$ (bullets), and $N=5\times 10^5$ (triangles). The solid curves are plotted using the dispersion relation, Eq.(\ref{KDR}), where $\omega_0$ and $k_0$ are determined from the numerical results and the core parameter $r_c=0.13$ $\mu$m for all cases. In contrast, the value of the healing length varies between 0.2 and 0.4 $\mu$m. The dashed curve is plotted for $N=1\times 10^4$ using $r_c=\xi=0.4$ $\mu$m highlighting the failure of approximating the core parameter $r_c$ by the healing length $\xi$. However, all cases shown in Fig. \ref{Fkelvons} are still in the Thomas-Fermi regime and for very small systems the dispersion relation shows much stronger finite-size effects. The modes below the continuum dispersion are the bending modes \cite{Simula2008b}. As the number of particles in the condensate is increased it becomes correspondingly longer and hence $k_0$ approaches zero.

\emph{Excitation method.---}
An elegant way to excite Kelvin waves has been realized in the experiment of Bretin \emph{et al.} \cite{Bretin2003a}. In this method, an elliptical rotating perturbation first populates the quadrupole $\ell=-2$ collective surface oscillation mode of the condensate thus reducing the orbital angular momentum of the system. Subsequently, the quadrupole mode population spontaneously decays pairwise into $\ell=-1$ Kelvin modes via the energy-momentum conserving Beliaev process in which a quadrupole quasiparticle interacts with the condensate, scattering to a pair of oppositely traveling Kelvin modes \cite{Bretin2003a,Mizushima2003a,Simula2008b}. Hydrodynamically one may explain such process in terms of a parametric (subharmonic) resonance between a quadrupole mode and a pair of Kelvin modes, leading to an elliptical instability of the initially straight vortex line \cite{Kerswell2002a,Simula2008b}. 

The shortcoming of this resonant excitation method is that it is constrained to a specific kelvon. However, it is possible to circumvent this constraint by deploying the core localized nature of the Kelvin modes \cite{Simula2008b}. By applying an external pinning potential, such as produced by a suitably detuned laser beam, at the vortex core thereby changing the local effective potential  it is possible to control the energies of the kelvons with respect to the condensate ground state energy. Since this may be performed in a fashion which does not dramatically affect the resonant quadrupole mode frequency, it is possible to modify and choose the kelvon wave vector which resonates with the quadrupole mode. Quantitatively such controlled kelvon excitation process may be achieved by using the time-dependent perturbation
\begin{eqnarray}
\label{KPERT}
V_{\rm pert}({\bf r},t)&=&V_0\frac{\sigma_0^2}{\sigma(z)^2}\exp(-2r^2 / \sigma(z)^2)  \\ 
&+& \epsilon m \omega^2_\perp ( \cos(2\Omega t)  ( x^2 - y^2) + 2 \sin(2\Omega t) xy )/ 2,  \notag
\end{eqnarray}
where the first line models a Gaussian laser field and provides the frequency shift to the kelvon dispersion. It is also worth noting that at finite temperatures the kelvon frequencies may experience an additional frequency shift \cite{Isoshima1999a,Simula2002a}. The weak potential, $\epsilon=0.025$, in the second line of Eq.(\ref{KPERT}) may be realized by rotating an elliptically deformed trap and is responsible for the resonant excitation of the quadrupole mode. In Eq.(\ref{KPERT}), $\Omega=-0.6\omega_\perp$ is the rotation frequency of the elliptical perturbation, $\sigma(z)=\sigma_0\sqrt{1+(z/z_{\rm R})^2}$ and we have chosen $\sigma_0=2\mu$m with a Rayleigh range of $z_{\rm R}=20\mu$m. It may be experimentally challenging to achieve focusing to such a narrow beam waist, however, the method also works for wider beams although the accuracy in determining the dispersion relation may then suffer. Furthermore, it would probably be wiser to switch on the pinning beam already before creating the condensate in order to avoid superfluous excitations. 

\begin{figure}
\begin{minipage}[t]{0.32\linewidth}
\includegraphics[width=\linewidth]{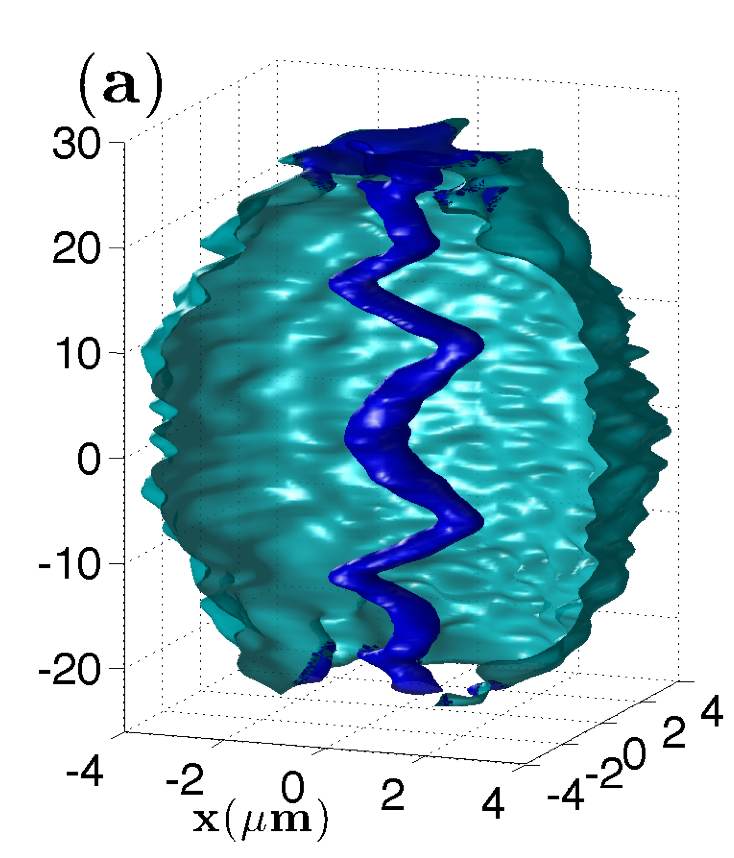}
\end{minipage}
\begin{minipage}[t]{0.32\linewidth}
\includegraphics[width=\linewidth]{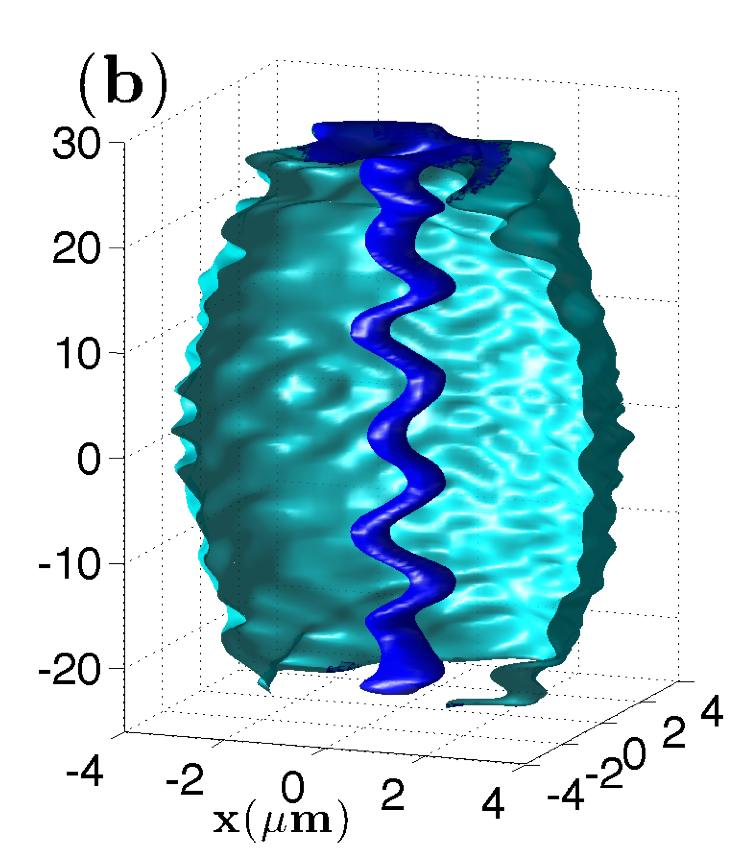}
\end{minipage}
\begin{minipage}[t]{0.32\linewidth}
\includegraphics[width=\linewidth]{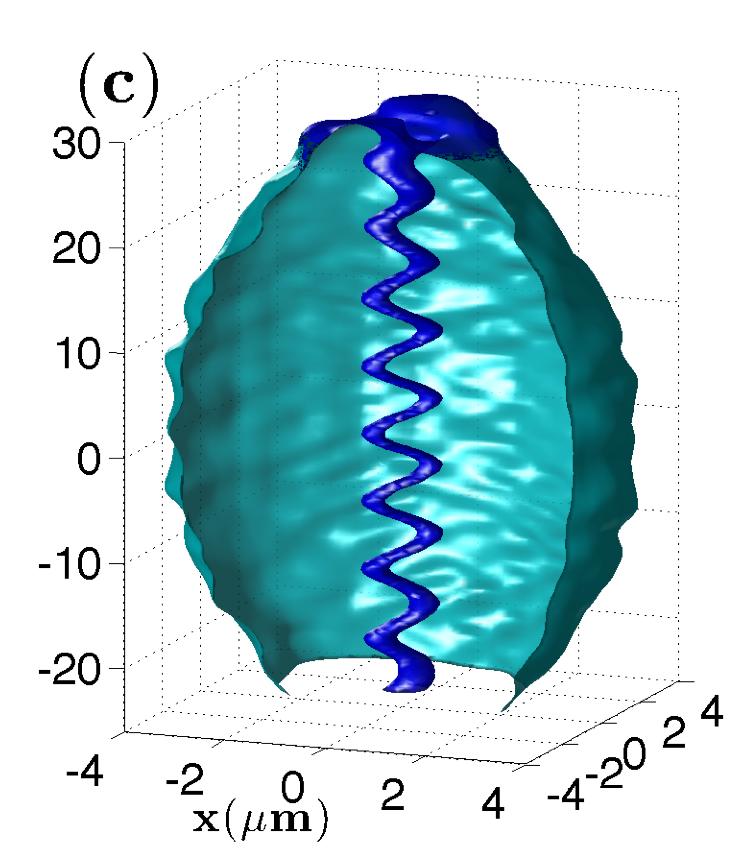}
\end{minipage}
\caption{Kelvin waves of different wave numbers excited using $V_0/\hbar\omega_\perp=6.0$ (a) 4.0 (b), and -4.0 (c). The duration of the quadrupole perturbation is 35 ms for all cases. The frames (a)-(c) are for times 75, 86, and 112 ms, respectively.}
\label{Fkelvin}
\end{figure}

Frames (a)-(c) of figure \ref{Fkelvin} show a gallery of kelvons with different wave vectors which have been generated by applying the above prescription with $V_0=\{6.0,4.0,-4.0\}\; \hbar\omega_\perp$, respectively. The duration of the perturbation was set to 35 ms and the plots are for times 75, 86, and 112 ms after the beginning of the perturbation. The initial in-plane sinusoidal shape of the vortex  
$ 
{\bf s_\pm}(\zeta,t)= 2R\cos(k_z \zeta)\cos(\omega t) {\bf\hat{e}_x} -2R\cos(k_z \zeta) \sin(\omega t) {\bf\hat{e}_y} +\zeta{\bf\hat{e}_z}
$
, seen in each frame of Fig.(\ref{Fkelvin}), forms as a superposition of two helical waves traveling in opposite axial directions.
It might also be possible to directly excite kelvons by using suitably tuned Laguerre-Gaussian beams or by applying an axially propagating velocity field such as a frequency matched traveling optical lattice with an appropriately chosen lattice period which could provide a resonant coupling to Kelvin modes.

\begin{figure}
\includegraphics[width=0.9\linewidth]{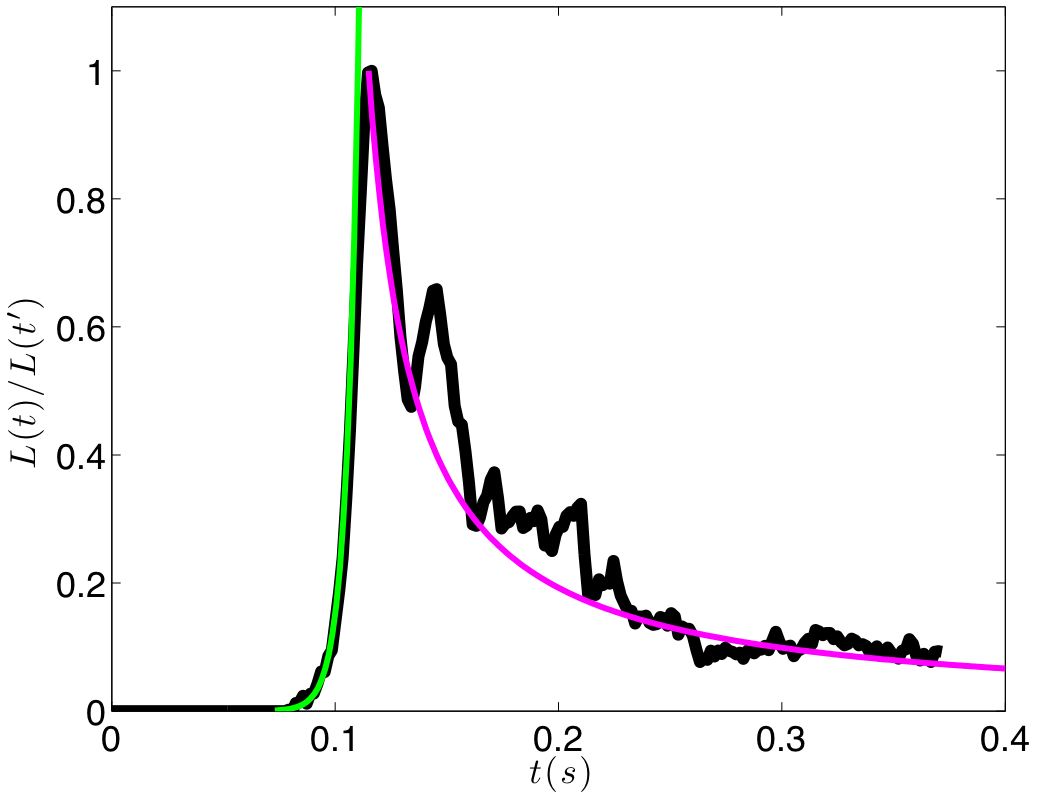}
\caption{Reduced length of the vortex as a function of time. Initially the vortex line remains straight until at 0.1 s its length rapidly increases due to the exponentially growing instability. The line length decay is modeled using the equation (\ref{linedecay}).}
\label{LL}
\end{figure}

\emph{Decay dynamics.---}
The decay of Kelvin waves may be studied by measuring the length of the vortex line 
\begin{equation}
S(t)=\int_{-Z}^{Z} \sqrt{1+[dX(z,t)/dz]^2 +[dY(z,t)/dz]^2}dz
\label{slength}
\end{equation}
as a function of time $t$, where $X(z,t)$ and $Y(z,t)$ are the parametrized $x$ and $y$ positions of the vortex core, respectively. We define a reduced line length density $L(t)=(S(t)/S(0)-1) m\omega_\perp/\hbar$, where $S(0)$ is the reference length of a straight vortex. The quantity $L(t)/L(t')$ is shown in Fig. (\ref{LL}), where $S(t')/S(0)=1.06$ and the data is obtained for $V_0/\hbar\omega_\perp=0$. At $t\approx0.1$ s the vortex line shape rapidly changes from being straight to oscillating sinusoidally in plane. The increase in the line length may be estimated by an exponential function, $A\exp(\gamma \omega_\perp t)$ plotted in Fig. (\ref{LL}) using $A=0.001$ and $\gamma = 0.3$. Between $t\approx0.1$ s and $t\approx0.3$ s the line length diminishes as the kelvons decay to kelvons of smaller wave numbers. The final state is a bent vortex which causes the data in Fig. (\ref{LL}) to saturate at a finite value. For our system, energy is conserved both during the growth and the decay of the line length. In the context of superfluid turbulence and Kelvin wave cascade, the decay of the vortex line length density is frequently modeled using a rate equation of the form \cite{Vinen1957a,Milliken1982a,Schwarz1988a,Leadbeater2003a,Walmsley2007a,Halperin2008a}
\begin{equation}
\frac{dL(t)}{dt}=-\nu L(t)^\alpha ,
\label{linerate}
\end{equation}
where, for $\alpha=2$, $\nu$ has the units of an effective kinematic viscosity. This may be integrated to yield
\begin{equation}
\frac{L(t)}{L(t')}=\frac{1}{1+\nu L(t') (t-t')}.
\label{linedecay}
\end{equation}
The decaying curve in Fig. (\ref{LL}) is plotted using $\nu/\kappa=0.2$, where $\kappa=2\pi\hbar/m$ is the quantum of circulation. It is interesting to compare this result to, e.g., those in Refs. \cite{Milliken1982a,Leadbeater2003a,Walmsley2007a}. However, more detailed investigation is required if one desires to draw a direct connection between the decay of Kelvin waves on a single vortex in a harmonically trapped Bose-Einstein condensate and the decay of vortex tangle in turbulent superfluid helium systems. 

\begin{figure}
\begin{minipage}[t]{0.49\linewidth}
\includegraphics[width=\linewidth]{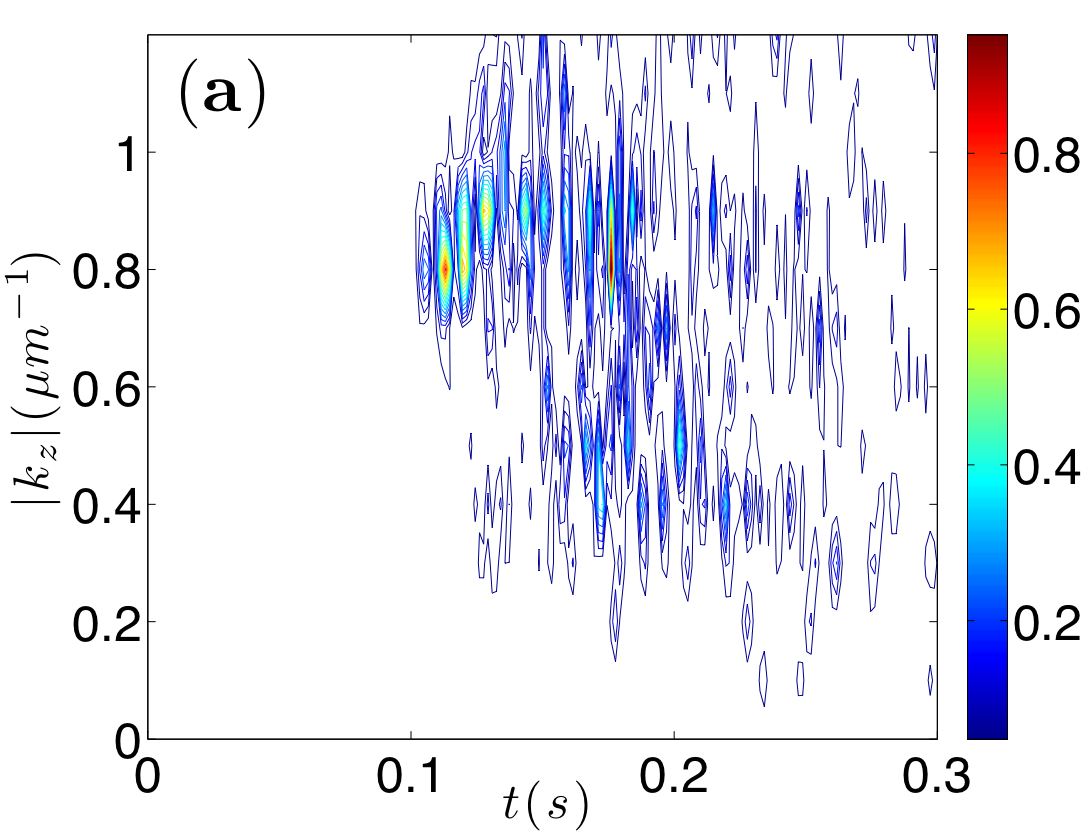}
\end{minipage}
\begin{minipage}[t]{0.49\linewidth}
\includegraphics[width=\linewidth]{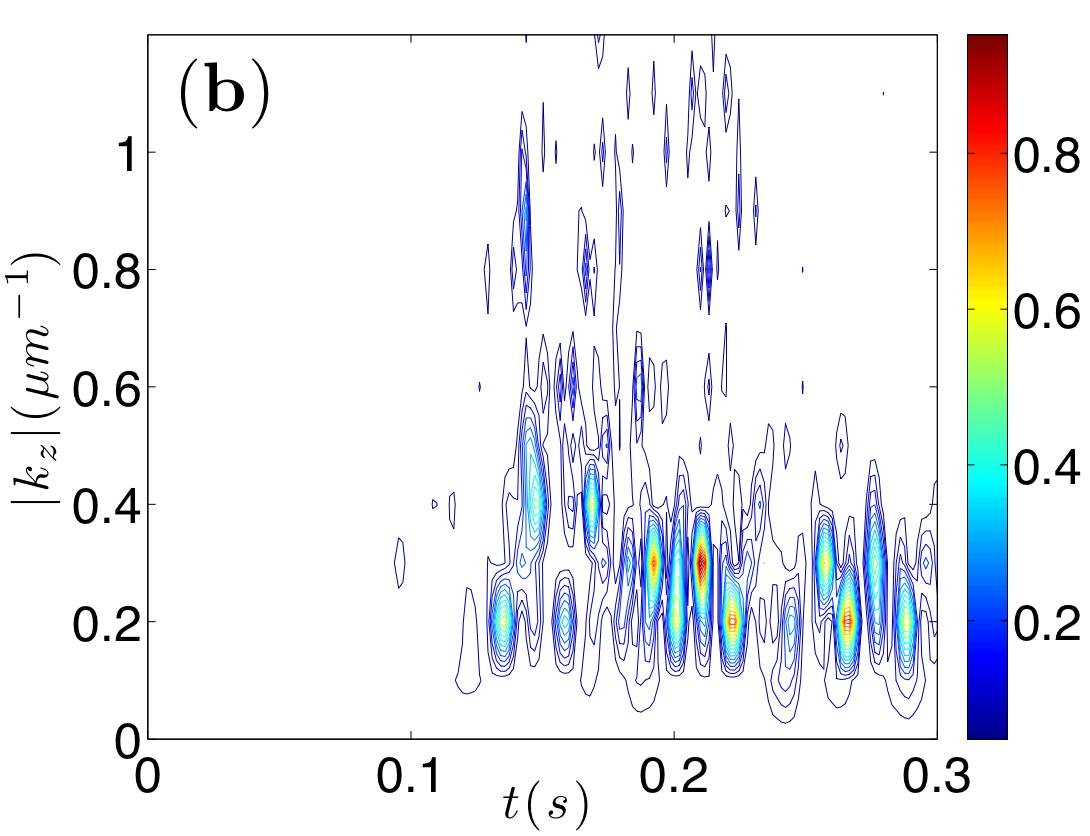}
\end{minipage}
\caption{Fourier transformed dipole moment $k^{1}_z$ (a) and monopole moment $k^{0}_z$ (b) of the system as a function of time.}
\label{Fmultipoles}
\end{figure}

Further insight to the excitation and decay of Kelvin waves may be obtained in terms of the multipole moments of the system. Initially, the condensate wavefunction attains strong quadrupole $\ell=-2$ moment since the elliptical rotating drive builds up population in the appropriate quadrupole mode. When the quadrupole mode population decays to $\ell=-1$ kelvons, the wavefunction gains a corresponding dipole moment and finally, as the kelvons further decay to phonons and kelvons of longer wavelength, they reveal themselves in the monopole moment of the wavefunction. Thus by analyzing the $z$-dependence of the real monopole $\ell=0$ and dipole $\ell=1$ moments
\begin{equation}
Q^\ell(z,t)=\int_0^\infty \int_0^\infty y^{\ell}|\psi(x,y,z)|^2 dxdy 
\label{moments}
\end{equation}
we may gain further information on the collective excitations present in the system. Figure \ref{Fmultipoles}(a) and (b) display the respective spatially Fourier transformed dipole $\ell=1$ and monopole $\ell=0$ signals 
\begin{equation}
k_z^\ell(t)=\mathcal{F}\{Q^{\ell}(z,t)-Q^{\ell}(z,0)\},
\label{ksignal}
\end{equation}
where $\mathcal{F}$ denotes a spatial Fourier transform, obtained for $V_0/\hbar\omega_\perp=0$. At $t\approx 0.1$ s a clear dipole moment signal emerges at $k^{1}_z=0.8$ and 0.9 $\mu$m$^{-1}$ signifying the onset of the kelvon excitation, see Fig. \ref{Fmultipoles}(a). Soon afterwards, a strong monopole signal emerges at $k^{0}_z=0.2$ and 0.4 $\mu$m$^{-1}$ due to the phonon emission from the decaying kelvons, see Fig. \ref{Fmultipoles}(b). Since the cylindrical symmetry is broken when the kelvons are excited, their presence in the system is manifest in both the monopole and dipole signals.

\subsection{Axisymmetric $\ell=0$ varicose waves}

\emph{Physical characterization.---} 
We assign the term varicose wave---or varicon to denote their quanta---to describe a class of axisymmetric, $\ell=0$, collective modes of the condensate. Here we focus on varicons of lowest frequencies which involve transverse motion and hence our varicose wave having zero axial momentum corresponds to the usual radial breathing (or monopole) mode \cite{Stringari1996a}. The modes with finite axial wave number exhibit similar sinusoidal breathing oscillation with $z-$dependent phase. In these systems the collective modes are density waves and therefore also the axial phonons produce similar density modulation along the vortex core due to the local density variations caused by the axially propagating sound waves \cite{Takeuchi2008a}. In contrast to the kelvons which only exist in the presence of a vortex and are localized in the vortex core, varicons correspond to the transverse swell-squeeze motion of the condensate and exist also in the absence of the vortex. 

\begin{figure}
\includegraphics[width=0.9\linewidth]{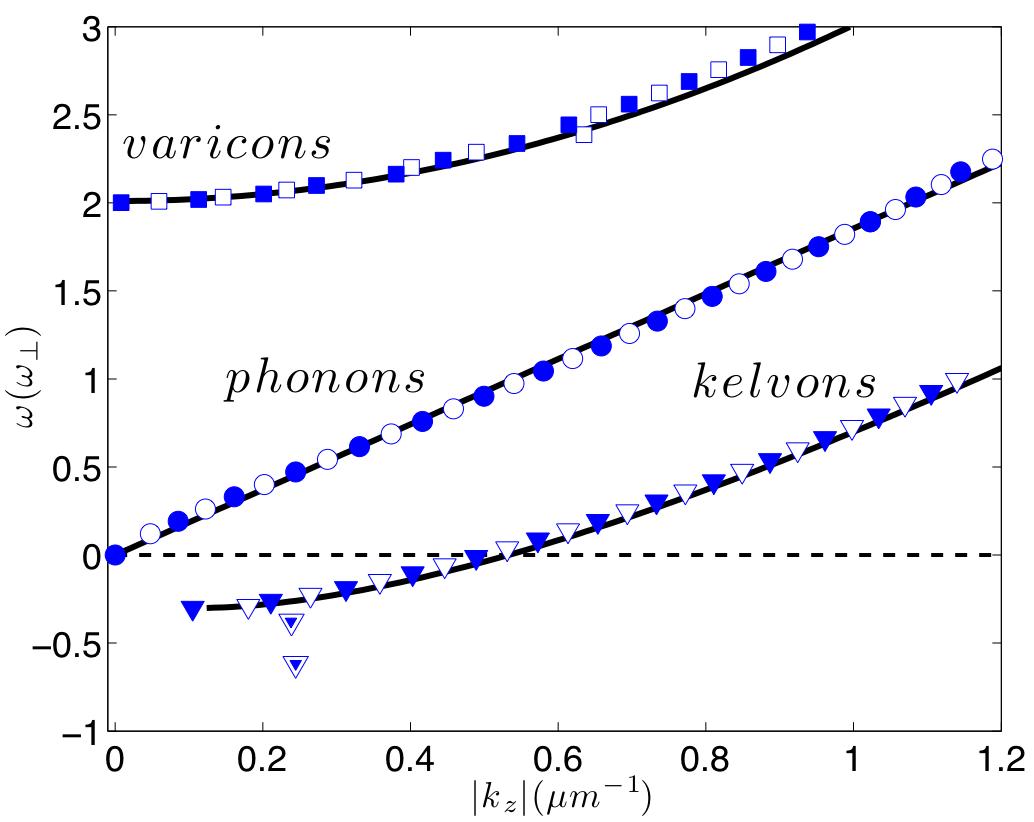}
\caption{Dispersion relation for varicose waves (boxes), sound waves (circles) and Kelvin waves (triangles). Solid (open) markers are plotted for even (odd) z-parity modes. The solid curves are the respective analytical varicon, phonon and kelvon dispersion relations discussed in the text.}
\label{FvaricoseDR}
\end{figure}

\emph{Dispersion relation.---}
Figure \ref{FvaricoseDR} shows the computed varicose mode frequencies (boxes) as a function of their wave vector. For comparison, we have also included both the phonon (circles) and kelvon (triangles) modes. The upper-most solid curve follows the frequency shifted free particle dispersion relation
\begin{equation}
\omega(k_z ) = \omega_0 +\frac{\hbar k^2_z}{2m}  
\label{VDR}
\end{equation}
describing a varicose mode in the presence of a vortex line. Here $\omega_0$ corresponds to the fundamental monopole frequency of the condensate \cite{Stringari1996a}. The lowest curve is the kelvon dispersion relation, Eq.(\ref{KDR}), and the straight line is the linear phonon dispersion $\omega_p=c|{\bf k}_z|$, where $c$ is the density averaged speed of sound. 

\begin{figure}
\begin{minipage}[t]{0.32\linewidth}
\includegraphics[width=\linewidth]{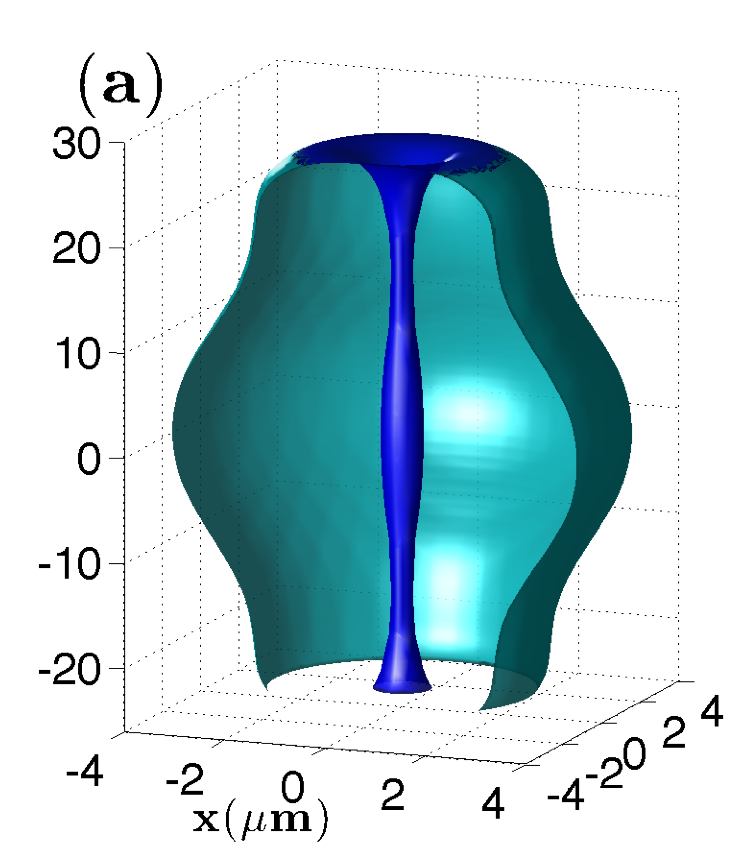}
\end{minipage}
\begin{minipage}[t]{0.32\linewidth}
\includegraphics[width=\linewidth]{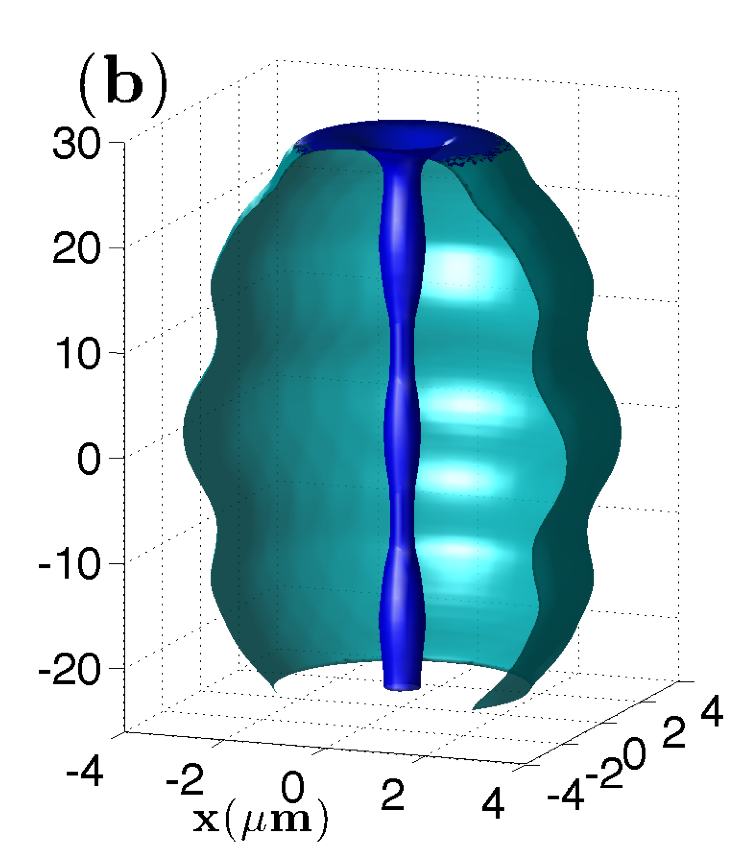}
\end{minipage}
\begin{minipage}[t]{0.32\linewidth}
\includegraphics[width=\linewidth]{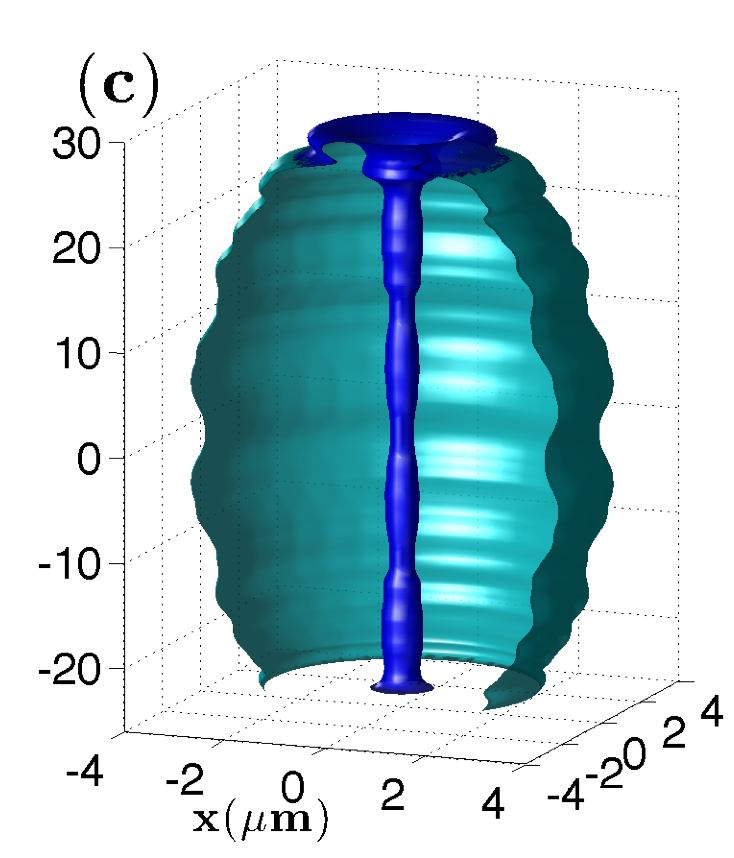}
\end{minipage}
\caption{Varicose waves of different wave vectors $k_z = 0.2$ (a) 0.4 (b), and 0.6 $\mu$m$^{-1}$(c). }
\label{Fvaricose}
\end{figure}

\emph{Excitation method.---}
Varicose waves may be resonantly excited using a perturbing potential of the form  
\begin{equation}
V_{\rm pert}({\bf r},t)=\epsilon m \omega^2_\perp ( x^2 + y^2) \cos(\Omega t) \cos(k_z  z),  
\label{VPERT}
\end{equation}
where $\Omega$ estimates the varicon frequency and $k_z$ is the corresponding wave vector. In the limit $k_z\to0$ the perturbation corresponds to a simple modulation of the strength of the radial trapping potential  resulting in the excitation of the fundamental monopole breathing mode. Figure \ref{Fvaricose} shows a collection of varicose waves of different wave numbers excited using the potential given by Eq.(\ref{VPERT}), where the frequency and wavevector are chosen according to Fig. \ref{FvaricoseDR}. The duration of the perturbation is 24 ms and the frames (a)-(c) are plotted for times 48, 26, and 27 ms, respectively.

\begin{figure}
\includegraphics[width=0.9\linewidth]{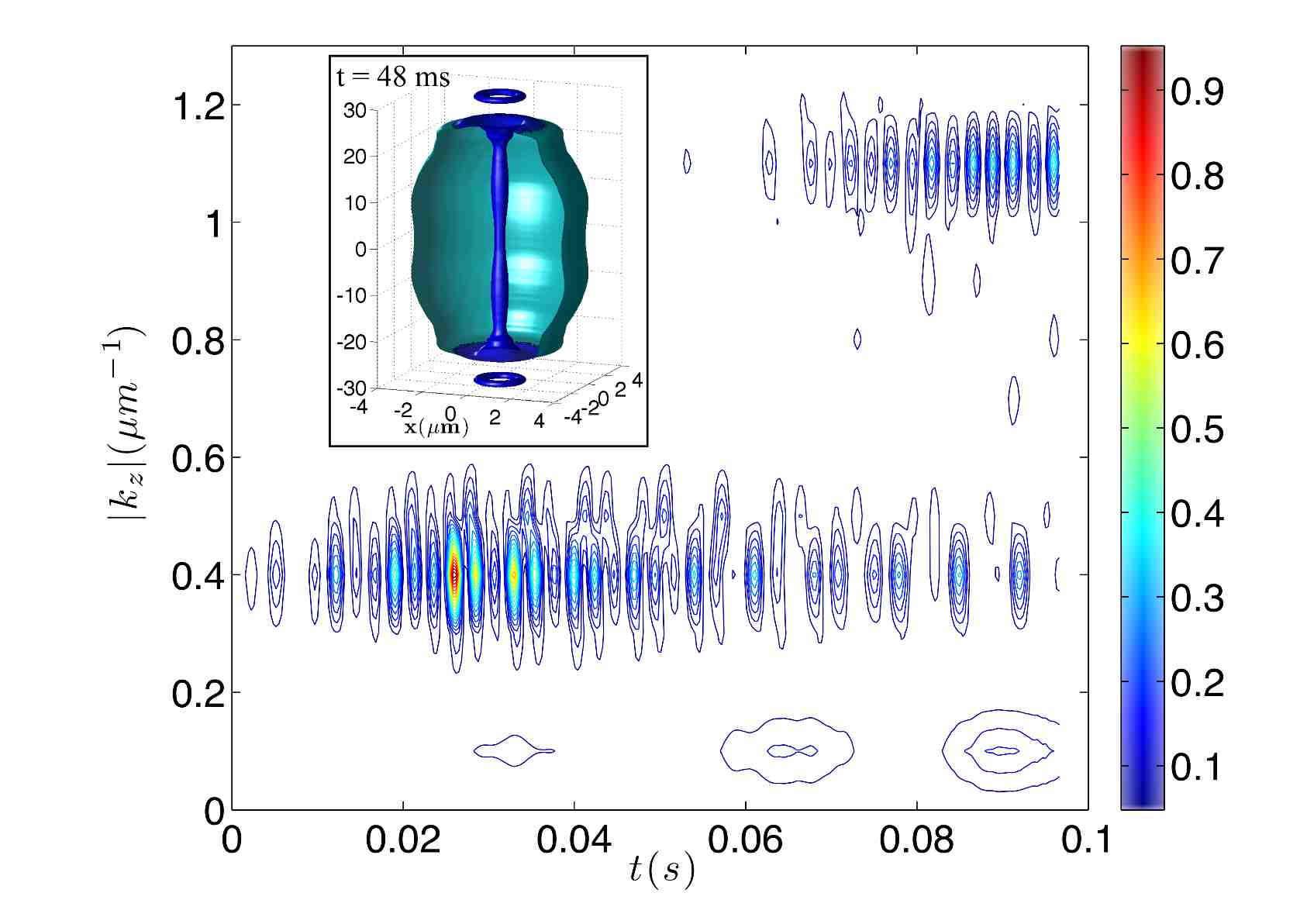}
\caption{Wave vector as measured from the monopole moment according to Eq.(\ref{ksignal}) as a function of time. The varicon with $k_z=0.4$ $\mu$m$^{-1}$ is first resonantly excited by the external perturbation. For later times, these varicons couple to the phonon branch, in particular to the one with $k_z=1.1$ $\mu$m$^{-1}$. The inset shows the condensate at $t=48$ ms.}
\label{Varidec}
\end{figure}

\emph{Decay dynamics.---}
Figure \ref{Varidec} displays the wave vectors of the varicose waves as measured from the monopole moment $Q^0$ according to Eq.(\ref{ksignal}), as a function of time.
The resonant external perturbation first excites the varicon with $k_z=0.4$ $\mu$m$^{-1}$ which then couples to phonons and in particular to the one with $k_z=1.1$ $\mu$m$^{-1}$, see also Fig.\ref{Varidec}. The varicose wave perturbations propagate in both directions along the vortex axis. On arriving at the low-density ends of the condensate, the varicose waves produce ring currents which become pinched off from the mother condensate as seen in the inset of Fig.\ref{Varidec}. 

\section{Discussion}
In conclusion, we have studied axially propagating collective excitations of quantized vortex lines in harmonically trapped elongated Bose-Einstein condensates. In particular, we have focused on two types of axial vortex waves: helical Kelvin waves and axisymmetric varicose waves. 
We have studied the excitation mechanisms, dispersion relations and decay processes of both types of collective excitation modes. We have employed efficient computational methods to solve: (i) the time-dependent Gross-Pitaevskii equation for the full three-dimensional quantum dynamics of the vortex motion and (ii) the Bogoliubov-de Gennes equations for the stationary axisymmetric single vortex eigenstates. The combination of these two tools provides us with a comprehensive description for this zero temperature system.

The Kelvin wave dispersion relation at the long wave length limit has both positive and negative energy eigenmodes with respect to the condensate energy and these have distinct physical manifestations: positive modes counter-rotate with respect to the unperturbed superflow while the negative modes are co-rotating. The kelvon frequencies may be engineered using external pinning potentials. In particular, it is possible to choose the sense of rotation of particular kelvon and to excite kelvons of specific wave numbers. Experimentally, Kelvin waves may be created indirectly by exciting the quadrupole mode of the system and allowing it to resonate with a pair of kelvons.  

The only dissipative mechanism operational in the zero temperature limit which may return a turbulent superfluid system to a laminar flow state is a reduction of the total length of the vortex lines. It is believed that in an isolated vortex the Kelvin wave cascade process is responsible for this transfer of the excess energy from small to large wave numbers and eventually phonon radiation is expected to absorb the surplus energy of the turbulence \cite{Svistunov1995a,Vinen2001a,Kozik2004a,Vinen2006a,L'vov2007a,Halperin2008a}. In contrast to this scenario---if applicable to our freely decaying single vortex case---here the cascade proceeds predominantly toward smaller wave numbers, i.e., toward longer wavelengths and smaller kelvon frequencies. 

In addition, we have studied axisymmetric excitations which exhibit a varicose shape. The long wavelength dispersion relation of the varicose waves studied here is quadratic in the wave vector. The varicose waves are readily visualized as periodic modulations of the vortex core diameter along the vortex axis, although the modes themselves exist independent of the presence of a vortex. The varicons are found to decay via coupling to phonon modes of higher wave vector. Detailed understanding of quantized vortex waves and their dynamics may shed light on open questions regarding various vortical systems including turbulent superfluids.

\begin{acknowledgments}
 This work was financially supported by the Japan Society for the Promotion of Science (JSPS).
\end{acknowledgments}

\end{document}